\documentclass[preprint,12pt]{elsarticle}

\usepackage{amssymb}

\biboptions{square,sort,comma,numbers}

\journal{Nuclear instruments and Methods A}

\begin{document}

\begin{frontmatter}



\title{Suitability of high-pressure xenon as scintillator for gamma ray spectroscopy}


\author{F. Resnati$^a$, U. Gendotti$^b$, R. Chandra$^b$, A. Curioni$^a$, G. Davatz$^b$, H. Frederich$^b$, A. Gendotti$^a$, L. Goeltl$^b$, R. Jebali$^b$, D. Murer$^b$, A. Rubbia$^a$}

\address{$^a$ ETH Zurich, Institute for Particle Physics, Zurich, Switzerland}
\address{$^b$ Arktis Radiation Detectors Ltd}

\begin{abstract}
In this paper we report the experimental study of high-pressure xenon used as a scintillator, in the context of developing a gamma ray detector. We 
measure a light yield near 2 photoelectrons per keV for xenon at 40~bar. Together with the light yield, we also measured an energy resolution of~$
\sim$9\% (FWHM) at 662 keV, dominated by the statistical fluctuations in the number of photoelectrons. 
\end{abstract}


\end{frontmatter}


\section{Introduction}
\label{Introduction}

Xenon detectors have been used and are in use for a number of applications, in particular for the detection of gamma rays (for a partial list see \cite
{Dmitrenko:2008} and references therein, \cite{Bolozdynya:1997vn, Curioni:2007rb, Aprile:2008ft, Carugno:1996ip}. For exhaustive 
review work see \cite{Aprile:2009dv} and \cite{Aprile:Wiley}). 
Xenon detectors are also preeminent in the field of Dark Matter searches and neutrinoless double beta decay (e.g.\ the XENON experiment 
\cite{Aprile:2011hi}, the LUX experiment \cite{Akerib:2012ak}, and the EXO experiment \cite{Auger:2012ar}).  
Xenon is an attractive material for gamma ray detection, in particular thanks to its high atomic number ($\mathrm{Z}=54$) and, as a consequence, large cross-section for photoelectric absorption. Xenon is also a rather dense material (the density of liquid xenon is 3.1 g/mL, while the density of 50 bar xenon at 293 K is 0.4 g/mL), which allows to build compact and efficient detectors. 
The scintillation light from Xe is at 175 nm, in the VUV, and it is therefore rather difficult to detect efficiently.
While a number of works presenting high-pressure Xe ionization chambers can be found in the literature, very few are available that discuss the 
properties of high-pressure Xe as a scintillator (known to these authors: \cite{Bolotnikov:1999}). Compared to a liquid Xe detector, a high-pressure Xe 
detector does not require cryogenics, and this simplification may be important for some applications. 
It has to be noted that the energy resolution achieved in Xe gas proportional chambers is significantly better than the one in liquid Xe 
\cite{Bolotnikov:1997}. It is therefore very interesting to compare the energy resolution obtained from primary scintillation in Xe gas to the one measured 
for liquid Xe.

In this work we have performed an experimental study of the properties of high-pressure Xe as a scintillator, in the context of developing a gamma ray 
detector for the detection of Special Nuclear Materials\footnote{Project MODES\_SNM, http://www.modes-snm.eu/.}. 
Our first goal has been to study experimentally the light yield and energy resolution from a high-pressure Xe tube, studying the dependence on 
thermodynamic conditions as well. 
A short description of the apparatus is given in Sec.~\ref{Experimental apparatus}, and the results are discussed in Sec.~\ref{Results}.

\section{Experimental apparatus}
\label{Experimental apparatus}

We have used a high-pressure tube (designed to withstand pressures up to 200 bar) filled with high purity xenon \cite{Chandra:2012}. 
The tube has a length of  200 mm and a diameter of 44 mm, with UV transparent windows at both ends. 
The inner surface of the tube is lined with a reflector combined with a wavelength shifter, 
that shifts the VUV light from xenon scintillation into visible light\footnote{The tube is produced by Arktis Radiation Detector Ltd. (http://www.arktis-detectors.com/)}.

\begin{figure}[htb]
\centering
\includegraphics[width=0.75\textwidth]{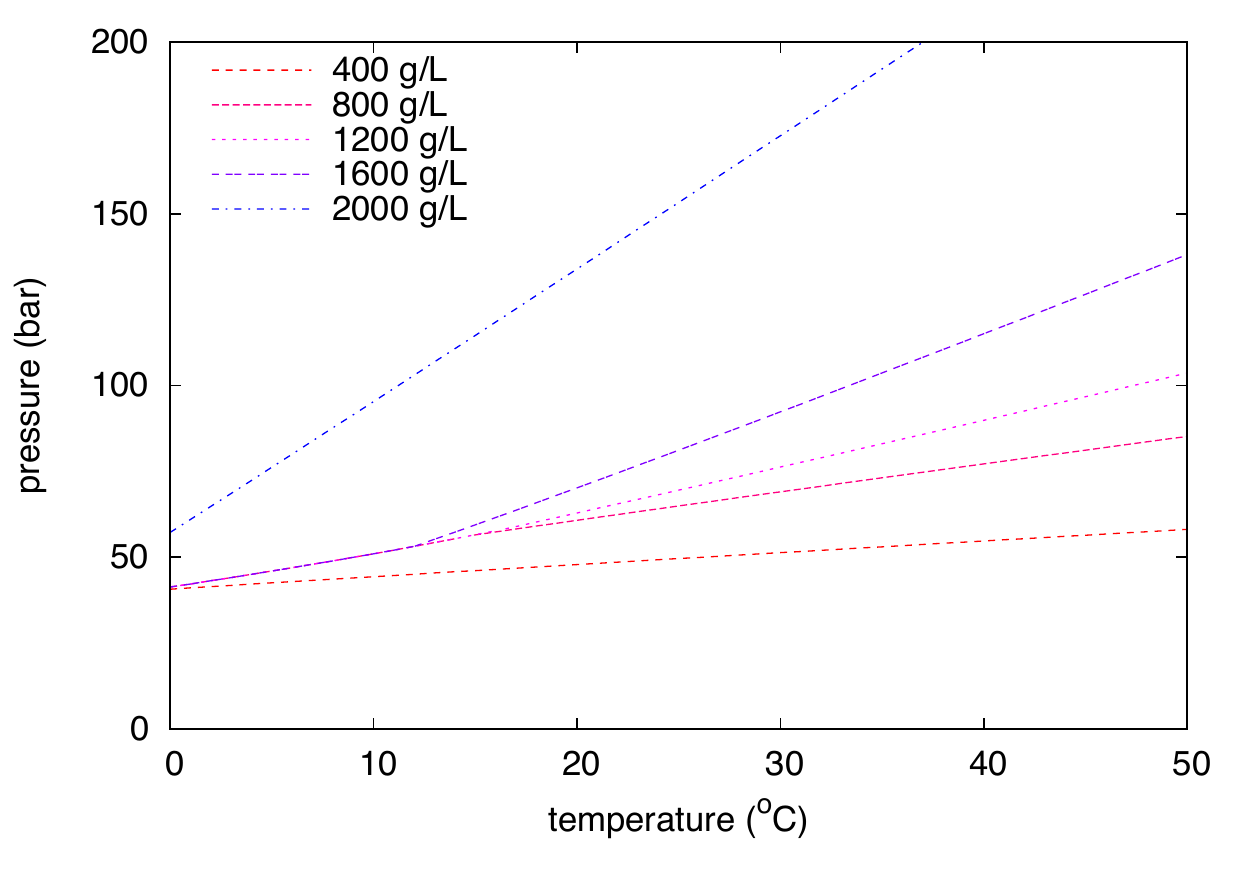}
\caption{ Isochoric curves for xenon. Data from NIST (http://webbook.nist.gov/chemistry/fluid/).}
\label{XePVsT}
\end{figure}

The isochoric curves of xenon are shown in Fig.~\ref{XePVsT}. The critical point for xenon is at 289.77~K and 58.41 bar, therefore at room temperature 
(293~K), xenon is in supercritical phase (for pressures above 58.41 bar), while for temperatures below 290~K it becomes liquid. 
The detector was operated in the pressure range between 30 and 60 bar, i.e.\  with xenon as a dense vapor (density changing between - roughly - 0.2 
and 0.7 g/mL).

\begin{figure}[htb]
\centering
\includegraphics[width=0.85\textwidth]{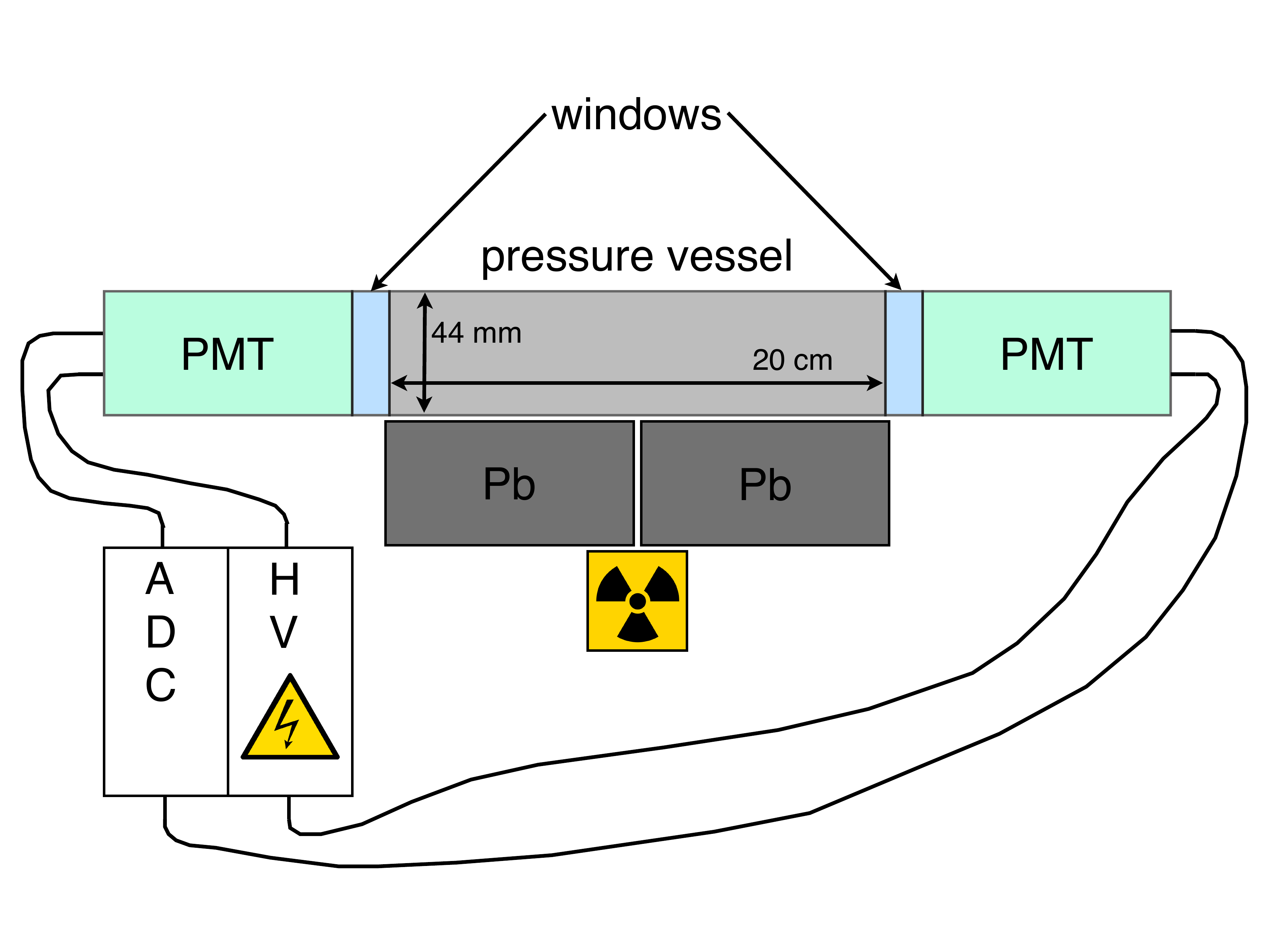}
\caption{ Schematic diagram of the experimental setup. See text for explanations.}
\label{schem}
\end{figure}

The scintillation light from the high-pressure tube was read out using two Hamamatsu R580 (quartz window)  photomultiplier tubes (PMTs). 
A schematic diagram of the experimental setup is shown in Fig.~\ref{schem}.
The data were taken using a CAEN V1751\footnote{http://www.caen.it/csite/CaenProd.jsp?parent=11\&idmod=602} digitizer (1 GS/s, 10 bit). 
A waveform is shown in Fig.~\ref{sampleEvent1}. 
The two PMTs were digitized independently, and the trigger, directly programmed on the digitizer, required  the coincidence of the two PMTs.

We took data with various gamma ray sources, most of the time with the source collimated with lead bricks, having a 3 mm wide slit illuminating the 
detector. 

\subsection{Analysis procedure}

\begin{figure}[htb]
\centering
\includegraphics[width=0.75\textwidth]{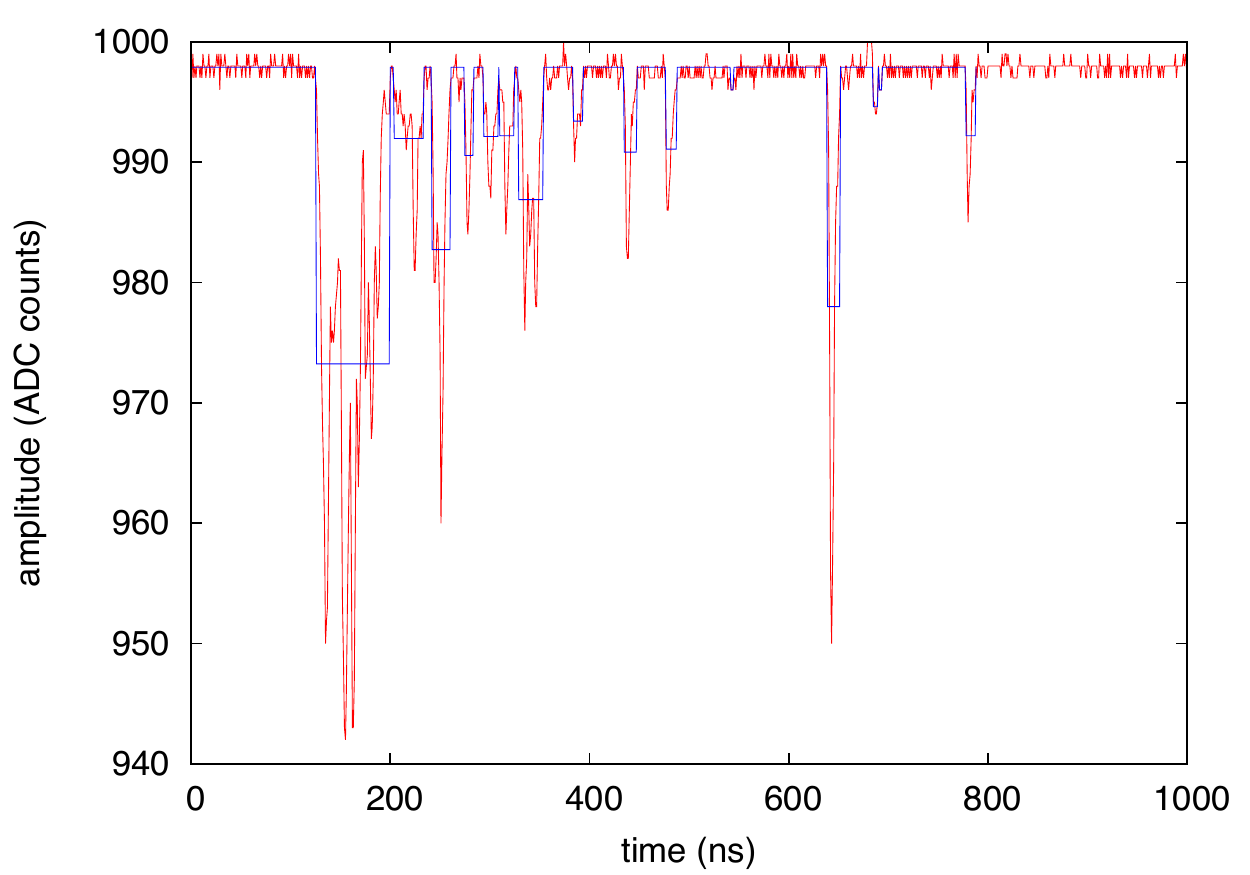}
\caption{ Digitized waveform from one the PMTs. The sampling rate is 1 GS/s, with a resolution of 10 bit. Also shown are the integration windows for the signal (see text for explanation).}
\label{sampleEvent1}
\end{figure}

\begin{figure}[htb]
\centering
\includegraphics[width=0.75\textwidth]{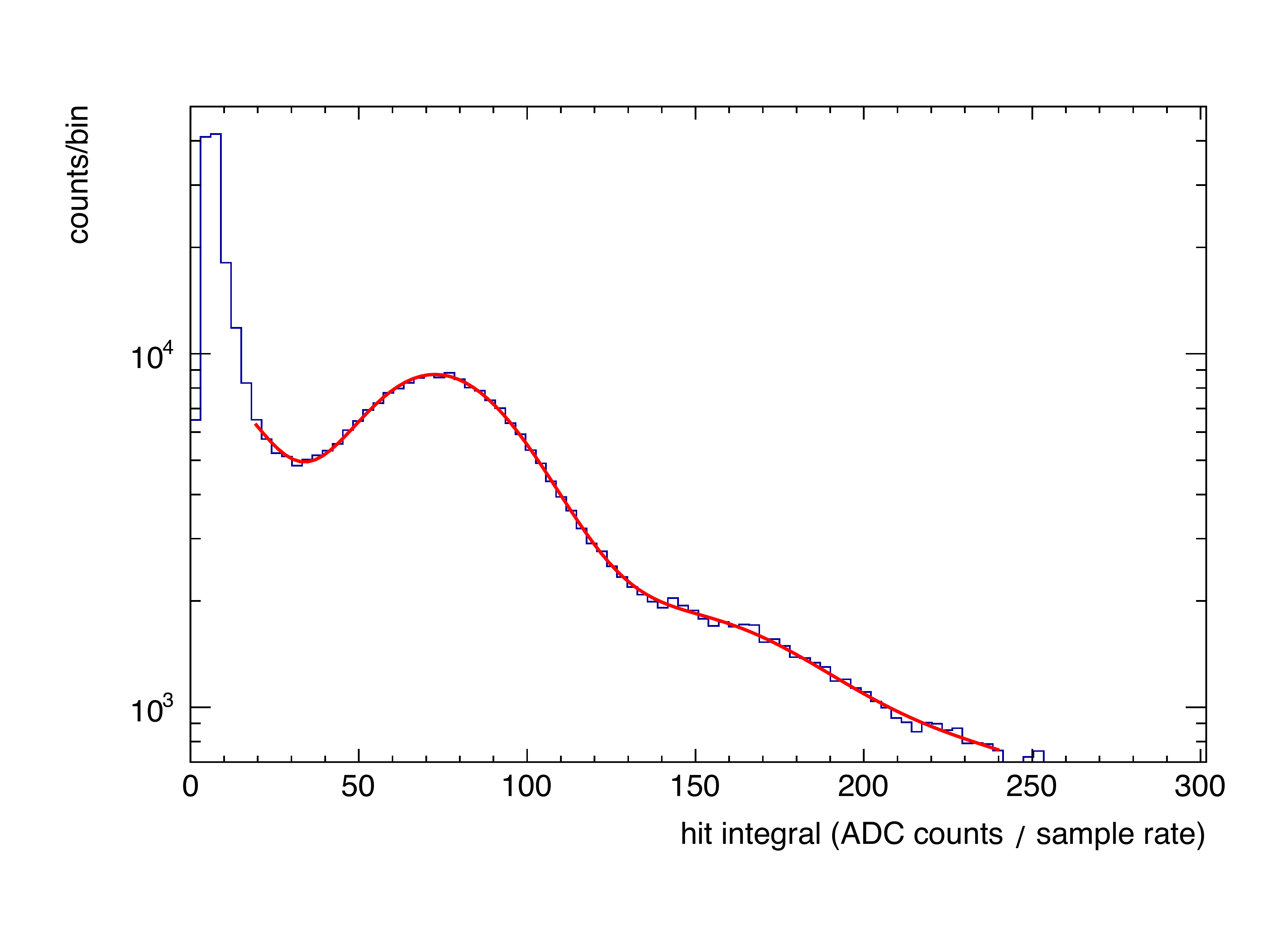}
\caption{ The single p.e.\  obtained from the digitized events. The superimposed function, fitted to the spectrum, includes a Gaussian to account for the 
background, and a sum of Gaussian functions to model the peaks of the $n$ p.e.\ ($n=1,2,...$).}
\label{singlePhotoelectronSpectrum}
\end{figure}

\begin{figure}[htb]
\centering
\includegraphics[width=0.75\textwidth]{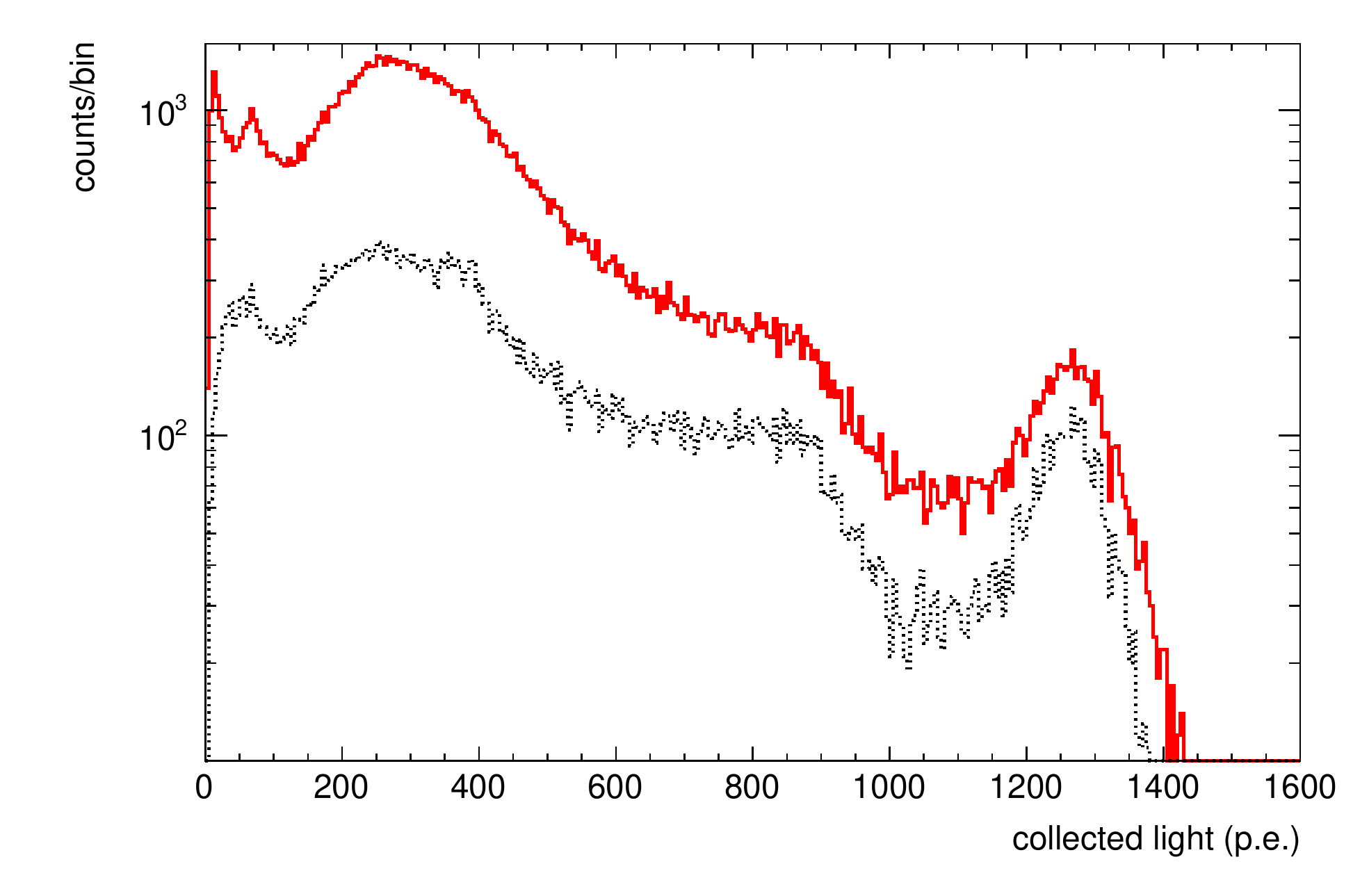}
\caption{ Energy spectrum of a collimated $^{137}$Cs before (continuous red line) and after (dotted black line) the selections described in the text.}
\label{137CsSpectrumLog}
\end{figure}

Our analysis procedure consists of four steps:
\begin{enumerate}
\item {\it Single photoelectron (p.e.) calibration:} The single p.e.\ is taken from the digitized events. A peak finding algorithm is applied to the digitized waveform (Fig.~\ref
{sampleEvent1}) and the area for each peak (in units of ADC counts$\times$ns) is determined integrating the signal over the time window in 
which the waveform is above threshold; the threshold is 2 times the noise (rms of the baseline), and a minimum of 3 consecutive samples above threshold is required. 
The time windows for each peak are shown in Fig.~\ref{sampleEvent1}.
The interesting region for the single p.e.\ is easily determined, producing a spectrum as the one shown in Fig.~\ref{singlePhotoelectronSpectrum}. The 
superimposed function, fitted to the spectrum, is
$$
A_0 g(0,s)+\sum_{n=1}^{n=3} A_n g(n \times m, \sqrt{n} \times \sigma),
$$
where $A_0 g(0,s)$ is a Gaussian with amplitude $A_0$, mean zero and sigma $s$ to account for the background, and the sum of the Gaussian functions models the photoelectron peaks.
\item {\it Each waveform is analyzed:} The procedure is similar to the one used for the single p.e. calibration, but this time the amplitudes of the peaks 
in the event are summed up, to define the number of p.e.\ per PMT for the given event. This integration procedure is performed over a time interval of 400 ns following the first threshold crossing. 
\item At this point it is possible to apply basic selections to the reconstructed events. We have three different selections: first of all, the time difference between the first identified peaks of the two PMT signals must be less than 40 ns (this selection rejects less than 0.1\% of the 
events). Second, the difference between the integrals of the two PMT signals is required to be less than 200 p.e.. Third, the integrals of the two PMT signals are required to differ less than 30\%. The last two selections  on the asymmetry between the two PMTs tend to improve the peak-to-total ratio 
in the energy spectrum and are especially effective for data with the source collimated in the center of the detector, because events showing largely 
asymmetric signals between the two PMTs tend to be only partially contained. One example showing the impact of these selections is given in Fig.~\ref
{137CsSpectrumLog}, for a collimated $^{137}$Cs source: after applying the selections described here, the peak-to-total ratio\footnote{Here for the ``total" we integrate over the entire spectrum, including the X-ray lines.} goes from 3.8\% to 7\%.
\item The event integrals for each PMT are summed up, giving the amount of light for each event (in units of p.e.) and the final spectrum.
\end{enumerate}

\section{Results}
\label{Results}

In the development of a gamma ray detector, our main goal was the determination of the light yield of high-pressure xenon, when exposed to gamma 
rays. 

\begin{figure}[htb]
\centering
\includegraphics[width=\textwidth]{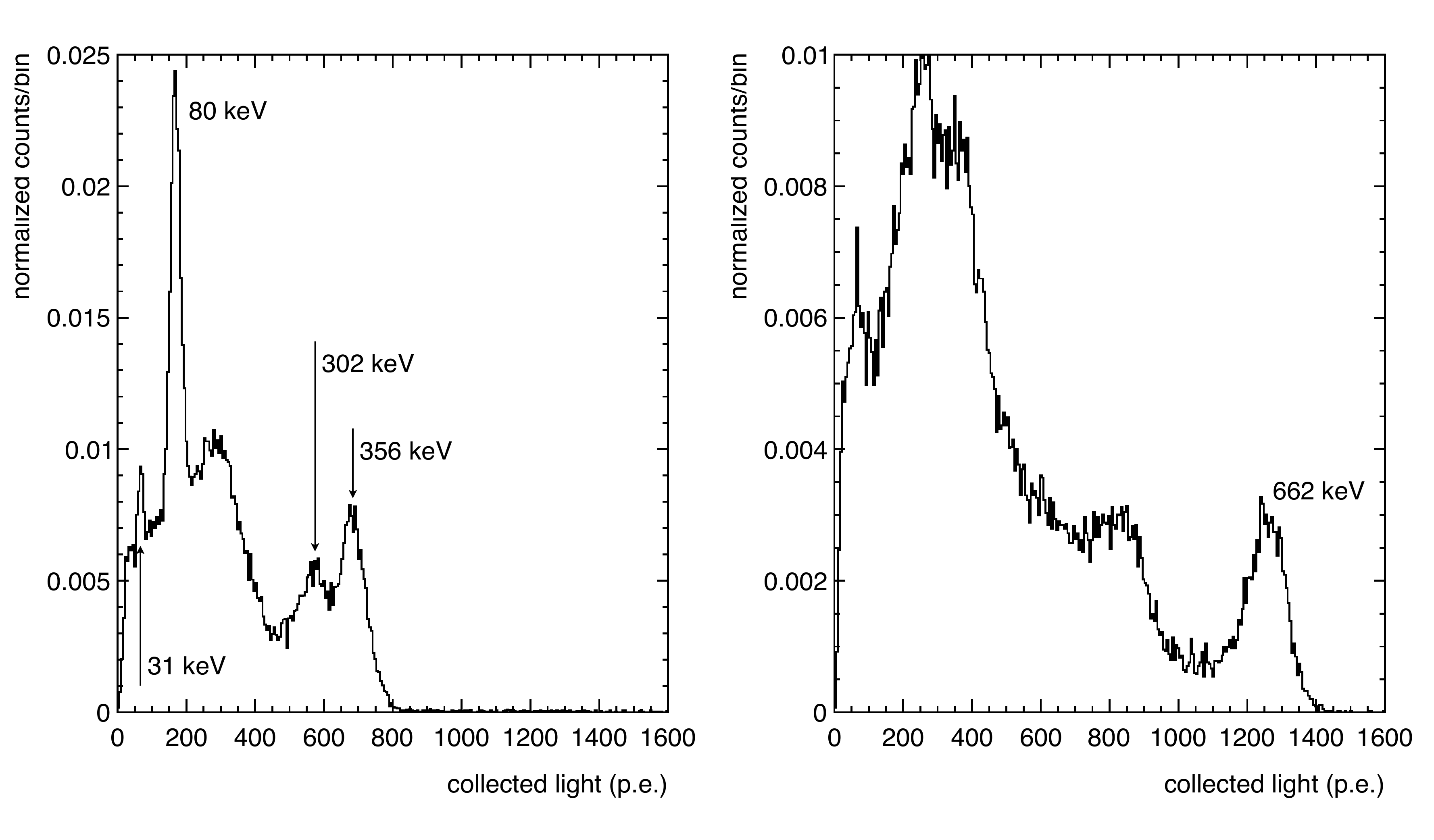}
\caption{{\it Left}: $^{133}$Ba spectrum. $^{133}$Ba gives lines at 31 keV (X-ray), 80 keV, 302 keV and 356 keV. At $\sim$30 keV several X-ray lines from iron and lead are also possible. {\it Right}: $^{137}$Cs spectrum, no off-line selections applied. Both sources were collimated in the center of the detector.}
\label{twoSpectra}
\end{figure}

The energy spectra from exposure to a $^{133}$Ba and a $^{137}$Cs are shown in Fig.~\ref{twoSpectra} (left and right, respectively).
In both cases the source was collimated at the center of the detector by lead bricks. 
$^{137}$Cs gives a line at 662 keV (and 32 keV X-ray line), while $^{133}$Ba gives several lines: the most prominent are at 31 keV (X-ray), 80 keV, 302 
keV and 356 keV, with additional weaker lines at 276 keV and 384 keV, which, with the available energy resolution, are partially merged with the 302 keV 
and 356 keV lines, and hard to detect individually. 
The backscattered peak is large for this particular detector geometry, where the mass of the stainless steel tube largely exceeds the mass of 
the xenon inside. 
Data were taken also with a $^{22}$Na source, which gives lines at 511 and 1275 keV.
From the data, we derive a light yield of $1.91 \pm 0.05$~p.e./keV at 662 keV.

\begin{figure}[htb]
\centering
\includegraphics[width=0.75\textwidth]{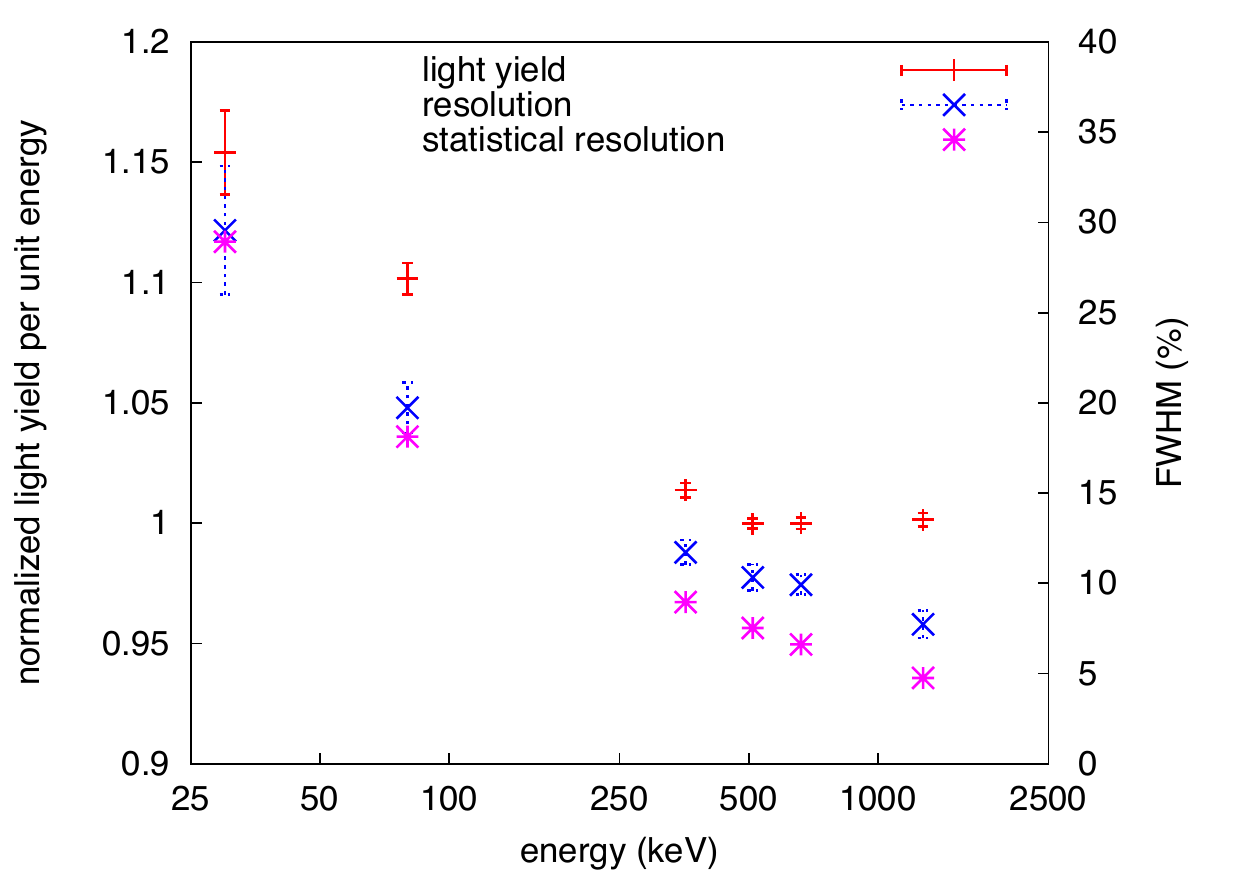}
\caption{ Normalized light yield, energy resolution, and limit to the energy resolution from photostatistics, $vs$ gamma ray energy. The normalized light yield is defined as the ratio between the light yield at a given energy divided by the light yield at 662 keV ($^{137}$Cs line). The energy resolution is derived from fitting the peaks in the spectrum with a Gaussian plus a second order polynomial for modeling the background. The ``statistical resolution" is calculated from the number of photoelectrons, assuming that the spread in the peak can be modeled with a Poisson distribution. }
\label{lightYeildAndResolutionVsEnergy}
\end{figure}

\begin{figure}[htb]
\centering
\includegraphics[width=0.7\textwidth]{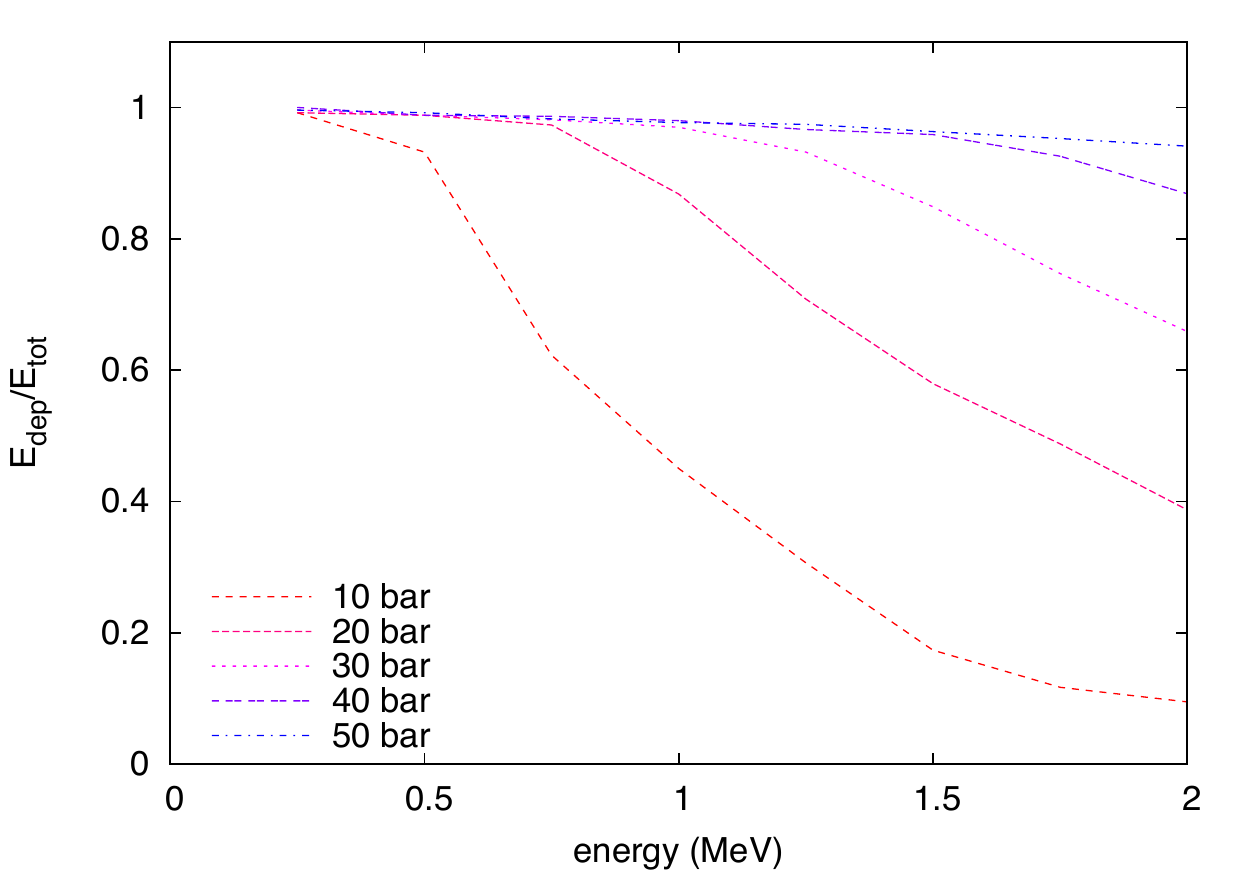}
\caption{ Fraction of energy deposited in a pressurized xenon cylinder of 20 mm radius for electrons starting at the center of the cylinder as a function of the electron initial energy, for  five different pressures. }
\label{e_range}
\end{figure}

The energy dependence of the light yield (per unit energy) is shown in Fig.~\ref{lightYeildAndResolutionVsEnergy}, where we present the ``normalized 
light yield", i.e.\ normalized to the light yield per keV at 662 keV ($^{137}$Cs). 
There is a systematic trend, showing a larger light yield at lower energies. 
We interpret this as due, at least in part, to non-containment of the electron from gamma conversion. Fig.~\ref{e_range}  shows the fraction of energy deposited in a pressurized xenon tube as a function of electron energy, for different gas pressures. Here monochromatic electrons (1000 for each energy) are generated in the center of a tube with a diameter of 40 mm, using the GEANT4 simulation package\footnote{http://geant4.cern.ch/}. At low pressures and densities the fraction of energy deposited drops rapidly with energy, because MeV electrons are, on average, not contained. At higher pressures the electrons stop inside the tube, but energetic Bremsstrahlung photons still escape from the tube. 
Considering the CSDA\footnote{Continuous Slowing Down Approximation, which gives an upper limit to the range of the electron. It can be easily calculated using the ESTAR program from NIST, http://physics.nist.gov/PhysRefData/Star/Text/ESTAR.html.} range for electrons, we see that it exceeds 20 mm in 40 bar xenon when the energy of the electron is larger than 1.1 MeV: given a diameter of the detector of 44 mm, the probability of full containment of electrons with a range exceeding 20 mm is rather small.

\begin{figure}[htb]
\centering
\includegraphics[width=0.75\textwidth]{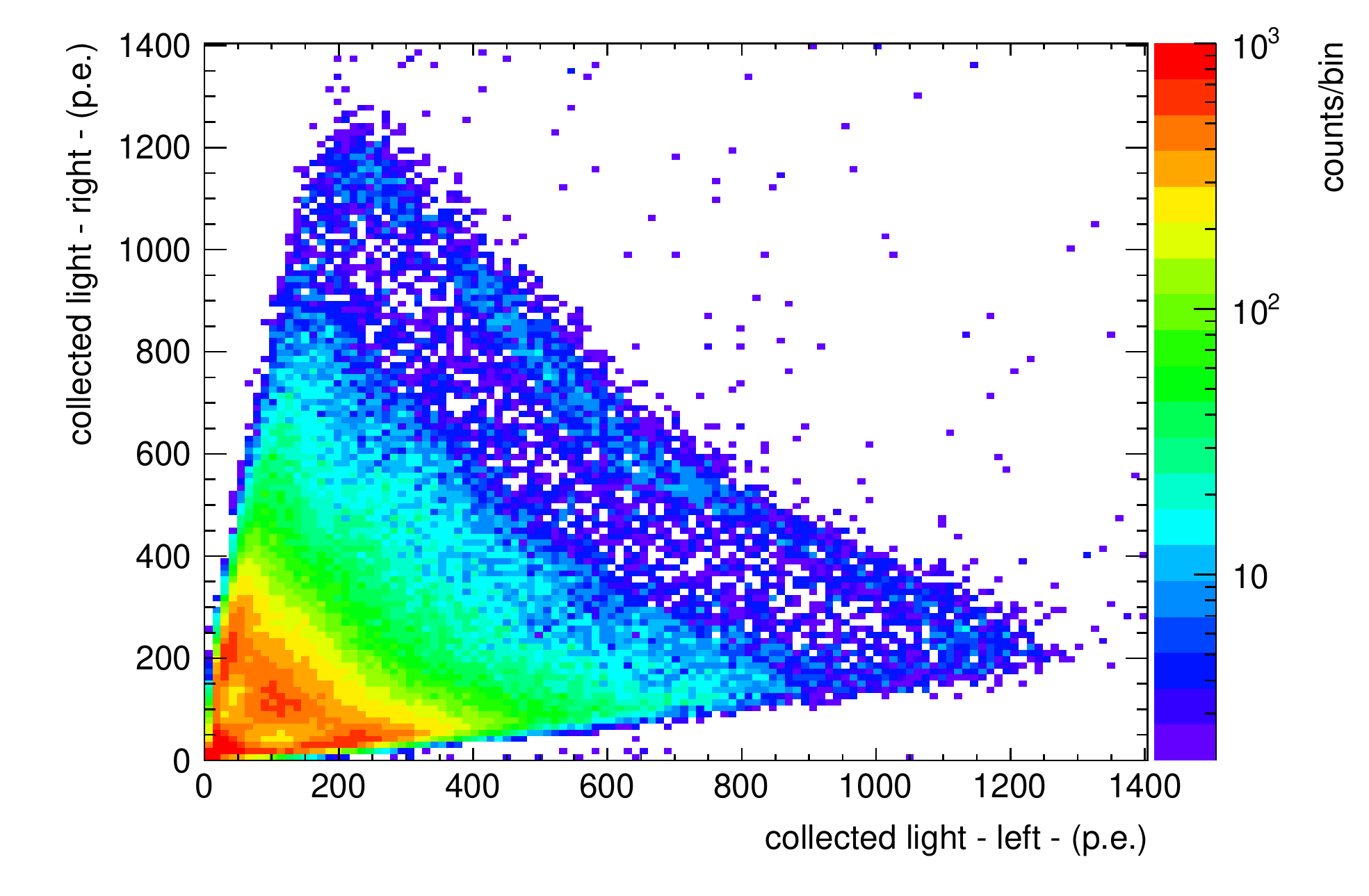}
\caption{ PMT-left vs PMT-right. Detector exposed to an uncollimated $^{137}$Cs source.}
\label{farVsNear}
\end{figure}

\begin{figure}[htb]
\centering
\includegraphics[width=0.7\textwidth]{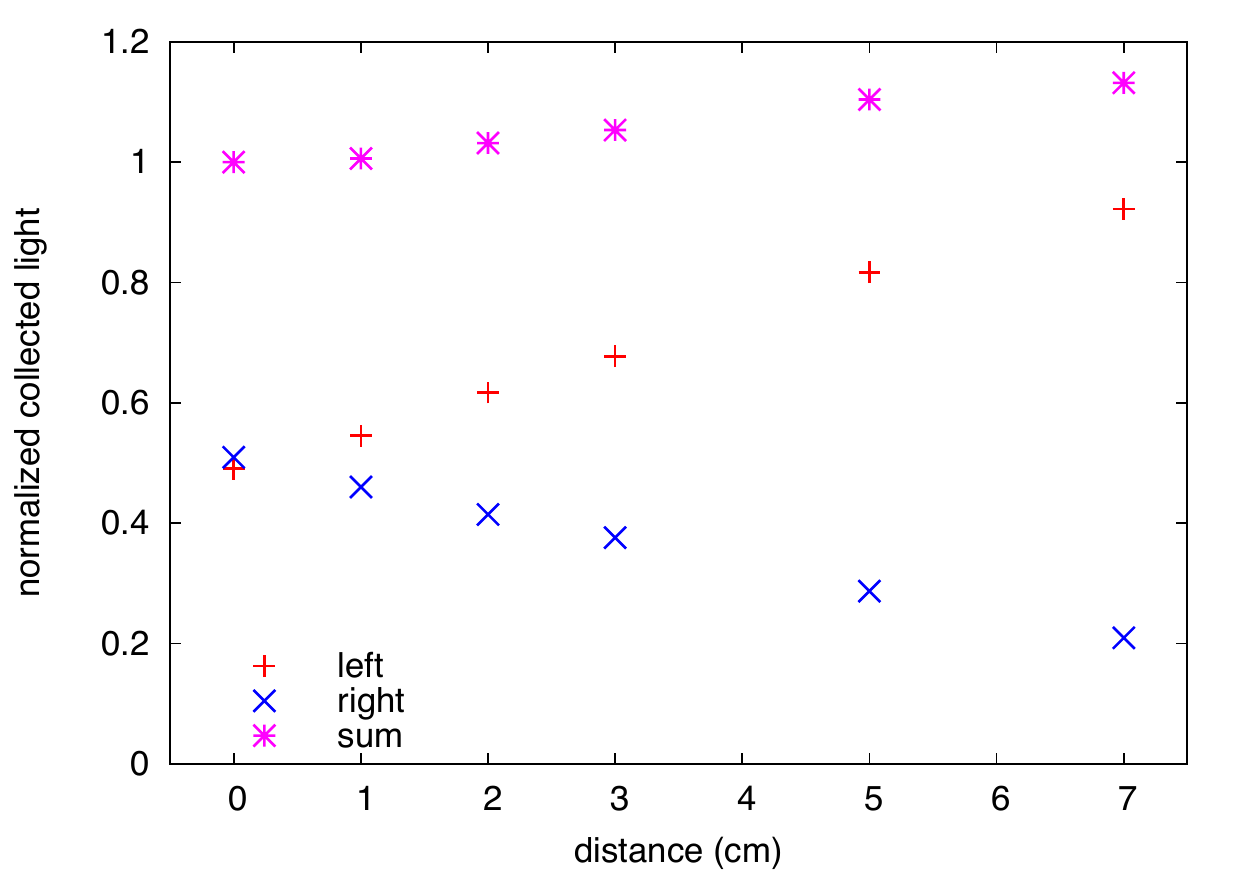}
\caption{ Collected light (normalized) as a function of the source position. Detector exposed to a collimated $^{137}$Cs source, moved along the detector.}
\label{positionScan-20}
\end{figure}

Again in Fig.~\ref{lightYeildAndResolutionVsEnergy}, the measured energy resolution and the expected energy resolution (for the measured light yield, 
and assuming that the spread can be modeled with a Poisson distribution), are shown. The measured energy resolution is consistently worse than the 
one expected from simple photostatistics. Again, the discrepancy increases with energy. The non-containment of the electron is one reason for this 
systematic effect. Another reason is that Compton scattering becomes more and more important when the energy of the gamma ray increases, and therefore the event is less localized and the energy is deposited within a larger fraction of the active volume. This leads to a degradation of the energy resolution, as can be seen in Fig.~\ref{farVsNear} and in Fig.~\ref{positionScan-20}, which show that in our setup the light yield has a significant (of the order of 15\% over the full volume) dependence on the interaction position inside the detector. 
In Fig.~\ref{farVsNear} we show the number of photoelectrons collected on one PMT vs. the one from the other PMT, for an uncollimated  $^{137}$Cs source at a distance of 2 m. 
It is clear that the band corresponding to the full energy peak has a ``banana shape", i.e.\ when the sharing of the light is almost equal between the two 
PMTs, the total collected light is less than when the light distribution between the two PMTs is very asymmetric.  
The data shown in Fig.~\ref{positionScan-20} were taken in different conditions, with a collimated $^{137}$Cs source, scanning the length of the tube up to 7~cm from the tube center. The dependence of the number of photoelectrons with respect to the position of the source is quite clear.
In fact, even when summing up both PMTs, we notice that the light yield for an event in the center of the detector is about 15\% lower than for an event 
happening next to one of the two PMTs. 
To summarize, the lower the energy of the gamma ray, the better the localization, because 
\begin{enumerate}
\item the range of the secondary electron is shorter;
\item the gamma ray is more likely to deposit all the energy in one single interaction (photoelectric absorption, to be contrasted with multiple Compton interactions).
\end{enumerate}
A better localization immediately translates in a reduced spread on the light yield, and on the energy resolution\footnote{A partial correction for the position dependence of the light yield can be introduced offline, allowing to recover the energy resolution.}.

With the help of a Monte Carlo simulation we estimate the light collection efficiency of 17\%, not including the quantum efficiency of the PMTs, with large 
systematic errors due to the lack of independent knowledge of the optical properties of the various materials. 
Given the geometry of the tube and an aspect ratio of almost 5, only a small fraction of the scintillation light is directly detected by the PMTs, without 
being converted after interaction on the detector walls. The light collection efficiency takes into account the propagation of the light in the tube, i.e.\ the 
wavelength shifting process on the detector walls (for VUV photons: assumed to be 100\% as estimated from available data), the reflection from the 
detector walls (for visible photons), the transmission/reflection from the windows (separately for VUV and visible photons), and the possibility of 
absorption.
The poor knowledge of the reflectivity of the material lining the tube introduces the largest uncertainty in our calculation. If we keep the 
reflectivity as the only free parameter in the calculation, we can determine it from the data, in particular from the position dependent light yield as shown in 
Fig.~\ref{positionScan-20}. We estimate a reflectivity of 93\% $\pm$ 2\%, which translates  into a light collection efficiency of $17 _{-4} ^{+5}$\%
The quantum efficiency of the PMTs is $\sim$27\%, both for visible and VUV photons. 
Therefore we measure an average energy expenditure (W$_\mathrm{s}$) of $26 _{-6} ^{+7}$ eV to produce one scintillation photon in high-pressure xenon at 662 keV. This can be compared to a value of 23.7 eV for liquid xenon \cite{Doke:1999ku}.

For completeness, we notice that our value for W$_\mathrm{s}$ is significantly smaller than the one of $76 \pm12$~eV from \cite{Parson:1990} for 15 bar Xe at 60 keV, and $72 \pm 6$~eV reported in \cite{Fernandes:2010gg} at much lower pressure, between 1 and 3 bar, at 5.9 keV. An even larger W$_\mathrm{s}$ of $111 \pm16$~eV was reported in \cite{doCarmo:2008} for 1 bar pressure at 5.9 keV. When comparing these conflicting results one should also consider the different thermodynamic conditions of the xenon.

\begin{figure}[htb]
\centering
\includegraphics[width=0.75\textwidth]{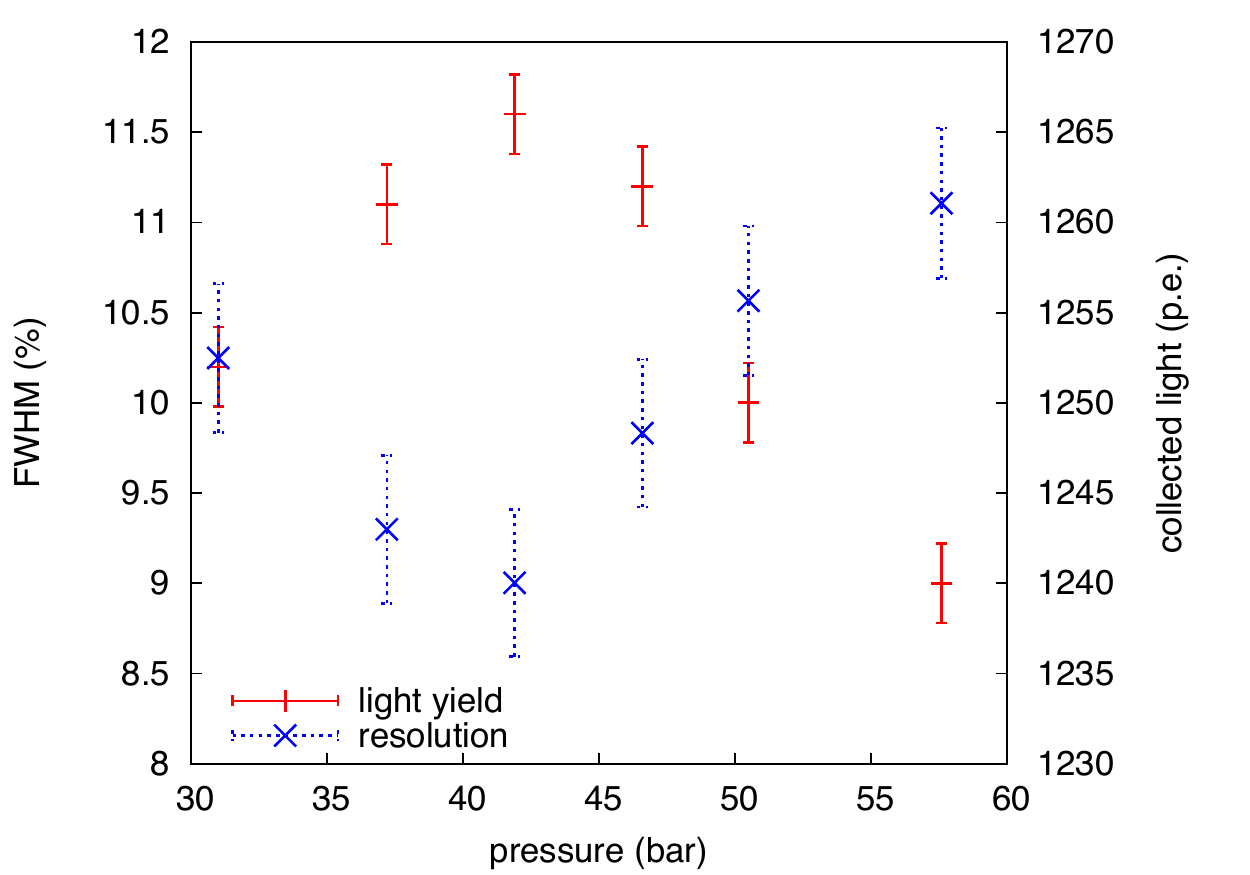}
\caption{ Light yield and energy resolution as a function of the xenon pressure. Detector exposed to a collimated $^{137}$Cs source.}
\label{lightYeildAndResolutionVsPressure-20}
\end{figure}

We have also studied the light yield as a function of the xenon pressure: results are shown in Fig.~\ref{lightYeildAndResolutionVsPressure-20}. From the 
point of view of an application, this is important because an efficient detector should be operated at the largest possible density. We clearly see that the 
energy resolution deteriorates when the xenon gas gets closer to the supercritical phase, essentially because many scattering  centers within the fluid 
hamper the uniform propagation of light. The abrupt change in the optical properties of the Xe is easily observable by simply watching the Xe cell while 
changing the pressure: the high pressure Xe goes from a perfectly transparent medium to a turbid one when the pressure exceeds 45 bars.
In fact, while the light yield does not change enough to modify much the Poisson term in the resolution, the effect is particularly visible on the measured 
energy resolution, pointing to a systematic smearing which applies to the light production or to the light propagation, or to both.

It is interesting to compare our results in terms of energy resolution (9\% FWHM at 662 keV, with a light yield of 1.9 p.e./keV) to the results obtained 
using liquid Xe detectors: Akerib et al. \cite{Akerib:2012ak} report a light yield of 8 p.e./keV and an energy resolution of 11.7\% FWHM at 662 keV, while Ni et al.  \cite{Ni:2006zp} report 6 p.e./ keV and 14.5\% (FWHM at 662 keV). Therefore we see that, while the energy resolution in high-pressure xenon is dominated by photostatistics, this is not the case for liquid xenon.

\section{Conclusion}
\label{Conclusions}

In this paper we have presented experimental results that give an initial characterization of the properties of high-pressure Xe gas as a scintillator, in 
particular in terms of the light yield. Using a detector which is not fully optimized for efficient light collection, we measure a light yield of $1.91 \pm 0.05$~p.e./keV at 662 keV at 40 bar pressure. The corresponding energy resolution is 9\% FWHM. 
We find this result encouraging for applications of high-pressure Xe as a scintillator for gamma ray spectroscopy. 

Some of the shortcomings of the present setup are well understood in terms of non-uniformity of the light yield and can be 
readily fixed in an optimized setup. A more uniform light collection across the active volume should push the energy resolution down to the limit given by 
photostatistics, which right now is reached only at the lowest energies.   
We have also studied the behavior of xenon as a function of the pressure, and we find that, when the gas gets close to its supercritical phase, the 
energy resolution degrades significantly, well beyond the effect of the slightly reduced light yield. This systematic effect is due to the presence of many 
scattering centers that affect the propagation of light in xenon, and determines an upper limit to the density of xenon used as a scintillator.

\section*{Acknowledgments}
This work is supported by the European Union through the MODES\_SNM project (Call FP7-SEC-2011-1, ERC grant agreement n$^{\circ}$ 284842).


\begin{thebibliography}{99}


 
\bibitem{Dmitrenko:2008}
V.V. Dmitrenko, S.E. Ulin, V.M. Grachev, K.F. Vlasik, Z.M. Uteshev, I.V. Chernyseva, K.V. Krivova, and A.G. Dukhvalov. Perspectives of High Pressure Xenon Gamma Ray Spectrometers to Detect and Identify Radioactive and Fissile Materials. S. Apikyan et al. (eds.), Prevention, Detection and Response to Nuclear and Radiological Threats, 155Ð175, 2008 Springer. (doi: \verb+10.1007/978-1-4020-6658-0_14+) 

\bibitem{Bolozdynya:1997vn}
  A.~Bolozdynya, V.~Egorov, A.~Kuchenkov, G.~Safronov, G.~Smirnov, S.~Medved and V.~Morgunov,
  Nucl.\ Instrum.\ Meth.\ A {\bf 385} (1997) 225.

\bibitem{Curioni:2007rb}
  A.~Curioni, E.~Aprile, T.~Doke, K.~L.~Giboni, M.~Kobayashi and U.~G.~Oberlack,
  Nucl.\ Instrum.\ Meth.\ A {\bf 576} (2007) 350
  [physics/0702078 [PHYSICS]].

\bibitem{Aprile:2008ft} 
  E.~Aprile, A.~Curioni, K.~L.~Giboni, M.~Kobayashi, U.~G.~Oberlack and S.~Zhang,
  Nucl.\ Instrum.\ Meth.\ A {\bf 593}, 414 (2008)
  [arXiv:0805.0290 [physics.ins-det]].
  
\bibitem{Carugno:1996ip}
  G.~Carugno, E.~Conti, A.~T.~Meneguzzo, R.~Onofrio, U.~Beriotto, S.~De Biasia, M.~Nicoletto and R.~Pedrotta {\it et al.},
  Nucl.\ Instrum.\ Meth.\ A {\bf 376} (1996) 149.
  
\bibitem{Aprile:2009dv}
  E.~Aprile and T.~Doke,
  Rev.\ Mod.\ Phys.\  {\bf 82} (2010) 2053
  [arXiv:0910.4956 [physics.ins-det]].

\bibitem{Aprile:Wiley}
Aprile, Elena; Bolotnikov, Aleksey E.; Doke, Tadayoshi (2006). Noble Gas Detectors. Wiley-VCH. pp. 8Ð9. ISBN 3-527-60963-6.

\bibitem{Aprile:2011hi}
  E.~Aprile {\it et al.}  [XENON100 Collaboration],
  Phys.\ Rev.\ Lett.\  {\bf 107} (2011) 131302
  [arXiv:1104.2549 [astro-ph.CO]].
  
\bibitem{Akerib:2012ak}
  D.~S.~Akerib {\it et al.}  [LUX Collaboration],
  arXiv:1210.4569 [astro-ph.IM].

\bibitem{Auger:2012ar}
  M.~Auger {\it et al.}  [EXO Collaboration],
  Phys.\ Rev.\ Lett.\  {\bf 109} (2012) 032505
  [arXiv:1205.5608 [hep-ex]].

\bibitem{Bolotnikov:1999}
A. Bolotnikov, B. Ramsey,
Nucl.\ Instrum.\ Meth.\ A {\bf 428}, 391 (1999)

\bibitem{Bolotnikov:1997}
A. Bolotnikov, B. Ramsey,
Nucl.\ Instrum.\ Meth.\ A {\bf 396}, 360 (1997)

\bibitem{Chandra:2012}
R. Chandra et al., 
JINST 7 C03035 (2012)
doi:10.1088/1748-0221/7/03/C03035

\bibitem{Doke:1999ku}
  T.~Doke and K.~Masuda,
  Nucl.\ Instrum.\ Meth.\ A {\bf 420} (1999) 62.
  
\bibitem{Parson:1990}
A. Parsons et al., 
IEEE Trans. Nucl. Sci. 37 (1990) 541.

\bibitem{Fernandes:2010gg}
  L.~M.~P.~Fernandes, E.~D.~C.~Freitas, M.~Ball, J.~J.~Gomez-Cadenas, C.~M.~B.~Monteiro, N.~Yahlali, D.~Nygren and J.~M.~F.~d.~Santos,
  JINST {\bf 5} (2010) P09006
   [Erratum-ibid.\  {\bf 5} (2010) A12001]
  [arXiv:1009.2719 [astro-ph.IM]].

\bibitem{doCarmo:2008}
S.~J.~C.~do~Carmo et al., JINST {\bf 3} (2010) P07004 doi:10.1088/1748-0221/3/07/P07004

\bibitem{Ni:2006zp}
  K.~Ni, E.~Aprile, K.~L.~Giboni, P.~Majewski and M.~Yamashita,
  JINST {\bf 1} (2006) P09004
  [physics/0608034].


\end{thebibliography}
\end{document}